\documentclass[aps,pra,twocolumn,showpacs,superscriptaddress]{revtex4-1}
\usepackage{graphicx}
\usepackage{dcolumn}
\usepackage{bm}
\usepackage{color}
\usepackage{times}
\usepackage{amssymb}
\usepackage{amsmath}
\usepackage{epsfig}
\usepackage{epstopdf}
\usepackage{dsfont}
\usepackage{subfigure}
\usepackage[colorlinks=true,citecolor=blue, linkcolor=blue,urlcolor=blue]{hyperref}
\usepackage{bookmark}
\usepackage{color}

\definecolor{Green3}{rgb}{0.80,0.87,0.76}

\begin{document}
\renewcommand{\thefootnote}{\fnsymbol {footnote}}
	
	\title{\textbf{Quantumness and entropic uncertainty   for a pair of static Unruh-DeWitt detectors}}
	
	\author{Yu-Kun Zhang}
	\affiliation{School of Physics \& Optoelectronic Engineering, Anhui University, Hefei 230601,  People's Republic of China}

\author{Tariq Aziz}
	\affiliation{School of Physics \& Optoelectronic Engineering, Anhui University, Hefei 230601,  People's Republic of China}

 \author{Li-Juan Li}
	\affiliation{School of Physics \& Optoelectronic Engineering, Anhui University, Hefei 230601,  People's Republic of China}

   \author{Xue-Ke  Song}
   \affiliation{School of Physics \& Optoelectronic Engineering, Anhui University, Hefei 230601,  People's Republic of China}

	\author{Liu Ye}
	
	\affiliation{School of Physics \& Optoelectronic Engineering, Anhui University, Hefei 230601,  People's Republic of China}

	\author{Dong Wang}
	\email{dwang@ahu.edu.cn}
    \affiliation{School of Physics \& Optoelectronic Engineering, Anhui University, Hefei 230601,  People's Republic of China}

\begin{abstract}
     In this {study}, we {investigate} a pair of detectors operating in Minkowski space--time and analyze the {characteristics} of various quantum resources within this framework. {Specifically, we focus on examining the properties of Bell nonlocality, quantum coherence, the nonlocal advantage of quantum coherence (NAQC), and  measured uncertainty in  relation  to the energy ratio and the distance between the detectors.  Additionally, we examine how the initial states influence these quantum properties.} {Notably, our findings reveal that both} a larger energy ratio and {a greater separation between the detectors degrade} the system's quantumness. {Moreover}, we explore the evolution of {entropic uncertainty} and demonstrate its inverse correlation with both Bell nonlocality and coherence, highlighting the intricate interplay between these quantum resources. {These insights provide a deeper understanding} of quantumness in a relativistic framework and {may contribute to the ongoing discussion on the black hole} information paradox.
\end{abstract}
\date{\today}
\maketitle

\section{Introduction}

{As {an emerging} discipline, quantum information {science integrates} quantum physics and information theory, {leveraging} the former to {investigate} the {fundamental} nature of reality at low temperatures and {on} microcosmic {scales}, while the latter {provides a framework for} information processing. The core concept of quantum information theory is the qubit} \cite{PhysRevLett.70.1895,PhysRevLett.85.461}. Unlike classical bits, which exist {solely in the states 0 or 1}, qubits can exist in a superposition of $|0\rangle$ and $|1\rangle$ simultaneously. {This property enables} quantum computing {to offer potential} advantages over classical computing for certain {computational} tasks \cite{PhysRevLett.86.5188,RevModPhys.80.1083}.

Quantum nonlocality refers to the phenomenon {in which} two {spatially separated} entangled particles exhibit correlated behaviors, such that a {measurement} performed on one particle {instantaneously} influences the state of the other. This instantaneous effect {contradicts} the {principles} of local realism. To address this {issue}, Bell introduced the concept of Bell's inequality \cite{PhysicsPhysiqueFizika.1.195}, {demonstrating} that {its} violation provides evidence {for} the existence of {nonclassical} quantum correlations. {Building on} this foundation, researchers {later} developed the Clauser-Horne-Shimony-Holt (CHSH) inequality {to further explore and test quantum nonlocality} \cite{PhysRevLett.98.140402,PhysRevA.80.032112,PhysRevLett.106.130402,PhysRevLett.23.880}. {This demonstrates} the maximum achievable violation of {these} limits. These advancements have further refined the theoretical framework of quantum mechanics, {providing} deeper {insights into} quantum entanglement and its implications.

{Quantum resources, including entanglement and coherence \cite{PhysRevD.107.045001,PhysRevLett.77.1413,PhysRevLett.114.210401,PhysRevLett.113.140401}, are considered as crucial aspects in the region of quantum information processing.} Quantum coherence, which {arises} from quantum superposition, {plays a key role} in quantum mechanics. It has evolved {alongside advancements in} quantum optics and quantum computing \cite{PhysRevA.69.062320,PhysRevA.99.052303}, {with becoming} emerging as its most prominent application. {To quantify coherence, several promising methods have been proposed, such as the $l_{1}$-norm coherence \cite{PhysRevA.93.012110,PhysRevA.95.032307,PhysRevA.96.052336} and {the} relative entropy of coherence \cite{PhysRevLett.119.150405,PRXQuantum.5.030303}.} In classical physics, local realism is regarded as an objective principle; however, in quantum {physics}, this principle can be {violated}.

{Another important quantum resource is called as the steerability of local coherence, which is formulated as a game between Alice and Bob based on coherence complementarity relations \cite{PhysRevA.95.010301}. For a two-qubit state $\hat{\rho}_{AB}$, local measurements on subsystem $A$, followed by classical communication, allow the average coherence of subsystem $B$'s conditional state to exceed the single-qubit coherence limit in mutually unbiased bases. As a result, the conditional state of subsystem $B$ can achieve a nonlocal advantage of quantum coherence (NAQC).}

{In the study of quantum systems, {uncertainty serves as a fundamental} tool for evaluating {their} states. In quantum information theory, {the} entropic uncertainty relation is {frequently} used to   {depict   the system's uncertainty}. In recent years, {this} relation has been {extensively} explored across various fields \cite{PhysRevA.75.022319,PhysRevA.86.042105,Pramanik2016,PhysRevA.87.022314,PhysRevLett.110.020402,PhysRevA.89.022112,PhysRevA.110.062220}, and its connections with other quantum resources have been investigated \cite{Zozor_2014,PhysRevA.91.042133,PhysRevA.93.062123,Huang2018,PhysRevA.102.012206,Ming2020,PhysRevA.106.062219,PhysRevE.106.054107,Li2022,PhysRevA.104.062204,Li2021,PhysRevA.101.032101,PhysRevE.109.064103,WANG2024138876}, thereby deepening our understanding of the quantum world.}

On the other hand, general relativity {describes gravity as} a {geometric property of space time}, and many {of its theoretical predictions} have {been} gradually confirmed. Although quantum information and relativity originate from distinct {domains}, {namely}, the microscopic and macroscopic worlds, intriguing connections {exist} between the two fields. Notable examples include quantum field theory (QFT) \cite{PasqualeCalabrese_2004}, relativistic quantum entanglement \cite{PhysRevLett.95.120404,PhysRevLett.109.130502}, and {the black-hole} information paradox \cite{Almheiri2013,Hawking1975,Marolf_2017}. {Consequently}, the investigation of relativistic quantum information has {attracted growing} interest in recent years.

{To} explore relativistic quantum information, a pair of two-level atoms, known as the Unruh-DeWitt detectors \cite{Reznik2003,SUMMERS1985257,VALENTINI1991321,10.1063/1.527733,PhysRevA.71.042104}, {has been extensively} studied. In general, {such detector pairs are employed to investigate Bell nonlocality, coherence, the nonlocal advantage of quantum coherence, and entropic uncertainty in the presence of a vacuum scalar field.} In this {study}, we {assume} that the two detectors interact independently {with} the vacuum field they occupy. For simplicity, the interaction {is} modeled as a monopole {coupling}. To avoid complications arising from motion, we assume that both detectors remain static indefinitely and that their interaction with the field is permanent. {This} model is {particularly significant for understanding the black hole} information paradox, {as a} Minkowski observer is equivalent to a freely falling observer in black hole {space--time}. All results presented in this {study} are {derived using} second-order perturbative approximations.

The {remainder} of this paper is {structured} as follows. In Section II, we provide a detailed description of the model {employed} in this study. Section III {explores} various {aspects of} quantumness, including nonlocality, coherence, and uncertainty. {Finally}, we {conclude} the paper with {a summary of our findings}.

\section{Model}
{Detector} models are {commonly employed to extract relativistic quantum information}. We {consider} a pair of two-level detectors, {referred to as} Alice and Bob, {with} energy-level gaps {given by} $\Delta E_{j} = E_{1_{j}} - E_{0_{j}}$ where $j$ corresponds to $A$ and $B$. We {prepare} this pair of detectors in an entangled state, {which} can be expressed as
\begin{align}
    | \phi \rangle = \sin\theta |0_{A} 0_{B} \rangle + \cos\theta |1_{A} 1_{B} \rangle,
    \label{E1}
\end{align}
{where} $\sin\theta$ and $\cos\theta$ are state parameters; {and} $|0\rangle$ and $|1\rangle$ {denote} the {ground and excited states} of {the} detectors, respectively. The detectors {are} placed in $(3+1)$-dimensional Minkowski {space--time}, {with their} trajectories {defined} as
\begin{align}
    t_{A} = \tau_{A},   & \qquad t_{B} = \tau_{B}, \nonumber  \\
    \mathbf{X}_{A} = 0, & \qquad \mathbf{X}_{B} = \mathbf{d},
    \label{E2}
\end{align}
{where} $\tau_j$ {denotes} the proper time of the $j$-th detector, and $\mathbf{d}$ is a constant vector {representing} the distance between {the} two detectors. {Additionally}, the interaction {between} the detectors {and} the real scalar field $\phi\left(x\right)$ is {modeled as} a monopole {coupling, which} can be {expressed} as follows:
\begin{align}
    S_{int}=\sum_{i=A,B} \nu_{j} \int d\tau_{j} \kappa_{j} m_{j}\left(j\right) \phi \left(x_{j}\left(\tau_{j}\right)\right),
    \label{E3}
\end{align}
{where} $\nu_{i}$ is the coupling constant, and $m_{j}\left(\tau_{j}\right)$ {represents} the {monopole} operator of the $i$th {detector, which} can be expressed as:
\begin{align}
    m_{j}\left(\tau_{j}\right) = e^{i H_{j} \tau_{j}} \left( |0_{j}\rangle \langle 1_{j}| + |1_{j} \rangle \langle 0_{j}| \right) e^{-i H_{j} \tau_{j}}.
    \label{E4}
\end{align}
{For} the $j$-{th} detector, $H_{j}$ denotes {the} free Hamiltonian, {while the} switching function $\kappa_{j}$ {governs the interaction duration}. The initial composite {state of the detector-field system} can be {expressed} as $|\Psi\rangle = |0_{M} \rangle \otimes |\psi\rangle$, where $|0_{M}\rangle$ denotes the real scalar field {in the vacuum state} and the detector system in Minkowski {space--time}. {By tracing out} the {field's degrees of freedom}, we {obtain} the initial density matrix of the detector system as follows:
\begin{align}
    \hat{\rho}_{AB}\left(t_{0}\right) =
    \begin{pmatrix}
             \sin^2\theta    & 0 & 0 & \sin\theta\cos\theta \\
                 0           & 0 & 0 &           0          \\
                 0           & 0 & 0 &           0           \\
        \cos\theta\sin\theta & 0 & 0 & \cos^2\theta
    \end{pmatrix}.
    \label{E5}
\end{align}
{During} the interaction of {a} real scalar field, the evolution of the composite system can be written as {follows}:
\begin{align}
    \hat{\rho}_{AB}\left( t \right) & = \mathrm{Tr}_{\psi}\left( \hat{T} e^{i S_{int}} |\Psi\rangle \langle \Psi | \hat{T} e^{i S_{int}} \right) \nonumber \\
                              & = \begin{pmatrix}
                                      \rho_{11} & 0         & 0         & \rho_{14} \\
                                      0         & \rho_{22} & \rho_{23} & 0         \\
                                      0         & \rho_{32} & \rho_{33} & 0         \\
                                      \rho_{41} & 0         & 0         & \rho_{44}
                                  \end{pmatrix},
    \label{E6}
\end{align}
where $\hat{T}$ denotes the time-ordering operator. {The} explicit forms of the density matrix above are presented {in} Ref. \cite{PhysRevD.107.045001}.

{Because both detectors are identical, their energy gaps satisfy $\Delta E_{A}=\Delta E_{B} \equiv \Delta E$}, and {their} coupling constants satisfy $\nu_{A} = \nu_{B} \equiv \nu$. We {assume that the} interaction duration {is} an adiabatic process, {allowing} the switching function $\kappa_{j}$ {to be set} to unity. {Consequently}, the elements of the {evolved} density matrix $\hat{\rho}_{AB}$ {reduce} to the following forms:
\begin{align}
    \rho_{11} & = \cos^{2}\theta \left( 1 - \nu^{2} P_{A}^{''} - \nu^{2} P_{B}^{''} \right), \nonumber \\
    \rho_{22} & = \cos^{2}\theta \nu^{2} P_{B}^{''}, \nonumber                                        \\
    \rho_{33} & = \cos^{2}\theta \nu^{2} P_{A}^{''}, \nonumber                                        \\
    \rho_{44} & = \sin^{2}\theta, \nonumber                                                           \\
    \rho_{14} & = \sin\theta \cos\theta \left(1 - \nu^{2} M_{A} - \nu M_{B}\right), \nonumber              \\
    \rho_{23} & = \cos^{2}\theta \nu^{2} \Re_{AB}, \nonumber                                           \\
    \rho_{32} & = \cos^{2}\theta \nu^{2} \Re_{AB}^{\ast}, \nonumber                                    \\
    \rho_{41} & = \sin\theta \cos\theta \left(1 - \nu^{2} M_{A}^{\ast} - \nu^{2} M_{B}^{\ast}\right),
    \label{E7}
\end{align}
where
\begin{align}
    P_{j}^{''} \left(\Delta E\right)   & =  \int \int d\tau_{j} d\tau'_{j} e^{-i\Delta E\left(\tau_{j}-\tau'_{j}\right)} G_{W} \left(x_{j}',x_{j}\right), \nonumber       \\
    M_{j} \left(\Delta E\right)   & =  \int \int d\tau_{j} d\tau'_{j} e^{-i\Delta E\left(\tau_{j}-\tau'_{j}\right)} \Theta\left(\tau_{j}-\tau'_{j}\right) \nonumber \\
                                  & *  \left(G_{W}\left(x'_{j},x_{j}\right)+G_{W}\left(x_{j},x'_{j}\right)\right), \nonumber                                        \\
    \Re_{AB}\left(\Delta E\right) & =  \int \int d\tau_{A} d\tau'_{B} e^{i \Delta E \left(\tau'_{B} - \tau_{A}\right)} G_{W}\left(x_{B}',x_{A}\right),
    \label{E8}
\end{align}
{herein}, $G_{W}\left(x'_{j},x_{j}\right)$ {represents the} positive frequency Wightman function. In the {above functions}, $P_{j} \left( \Delta E \right)$ denotes the transition probability of the $j$-{th} $(j\in \{A, B\})$ detector from the ground state to the excited state, {which is entirely determined} by the spontaneous emission probability. {Additionally}, $M_{j}(\Delta E)$ is determined by the expectation value of the anticommutator of the external scalar field, {while} $\Re_{AB}(\Delta E)$ represents the interaction between the field and detectors. {Moreover}, $G_{W}\left(x'_{j}, x_{j}\right) + G_{W}\left(x_{j},x'_{j}\right)$ {the term} in second equation of {Eq}. (\ref{E8}) can be {computed as} $\langle0_{M} | \left\{ \phi\left(x'_{j}\right), \phi\left(x_{j}\right) \right\}| 0_{M} \rangle$.
By {integrating Eq. (\ref{E8}), we obtain}
\begin{align}
    P_{j}^{''}                     & = \frac{\delta\left(0\right)}{2c^{3}} \sqrt{\Delta F} \nonumber ,                                                                      \\
    \mathrm{Re} \left(M_{j}\right) & = \frac{\delta\left(0\right)}{4c^{3}} \sqrt{\Delta F} \nonumber ,                                                                      \\
    \Re_{AB}                       & = \frac{\delta\left(0\right)}{2c^{3}} \sqrt{\Delta F} \frac{\sin\left(\frac{d}{c} \sqrt{\Delta F}\right)}{\left(\frac{d}{c} \sqrt{\Delta F}\right)} ,
    \label{E9}
\end{align}
where {$ \Delta F =(\Delta E^{2} - \left(mc^{2}\right)^{2})/\left(mc^{2}\right)^{2}$ {represents} the energy ratio between {the} detector and {the} external field, {effectively capturing} the energy difference between them}; $\mathrm{Re}$ denotes the real part of {the} equation; $d$ represents the distance {between the} two detectors; and $c$ represents the velocity of light in Minkowski {space--time}. {Additionally}, $\delta\left(0\right)$ is the {Dirac delta} function, {arising from the} infinite time integration, {and can be} expressed as:
\begin{align}
    \delta \left(0\right) = \lim_{T \to \infty} \frac{1}{2\pi} \int_{-T/2}^{T/2} du.
    \label{E10}
\end{align}
{It is evident} that the Dirac delta function {diverges}.
{Generally, computations of the relevant quantities are more convenient when the function is convergent.} {Because} the $\delta\left(0\right)$ function is independent of the other components except for time $T$, we can {evaluate} \cite{PhysRevD.107.045001} the quantum resources considered here by
\begin{align}
    \dot{\xi} & = \frac{\xi}{T} = \left(\mathrm{finite \ quantity}\right) \times \frac{\delta\left(0\right)}{T} \nonumber \\
              & = \left(\mathrm{finite \ quantity}\right) \times \lim_{T \to \infty} \frac{1}{2\pi T} \int_{-T/2}^{T/2} du,
              \label{E11}
\end{align}
where $\xi$ {represents} the quantum resources {under consideration}, including Bell nonlocality, coherence, {NAQC}, quantum uncertainty, and the integral in Eq. (\ref{E11}) can be calculated {using} the result for $\frac{1}{2 \pi}$.

\section{Quantum Correlation and Entropic Uncertainty}
{In this section, we first review Bell nonlocality and explore its dynamical evolution within the proposed model. Subsequently, we analyze two types of quantum coherence, namely, $l_{1}$-norm coherence and relative entropy coherence, demonstrating how the system's coherence evolves over time. Building on these measures, we investigate the nonlocal advantage of quantum coherence, assess state properties through this nonlocal advantage, and compare the characteristics of the nonlocal advantages of the two types of coherence. Finally, we explore quantum uncertainty within this framework and reveal its relationship with the aforementioned quantum resources.}

\subsection{Bell nonlocality}
\begin{figure}[t]
    \centering
   \includegraphics[width=1\linewidth]{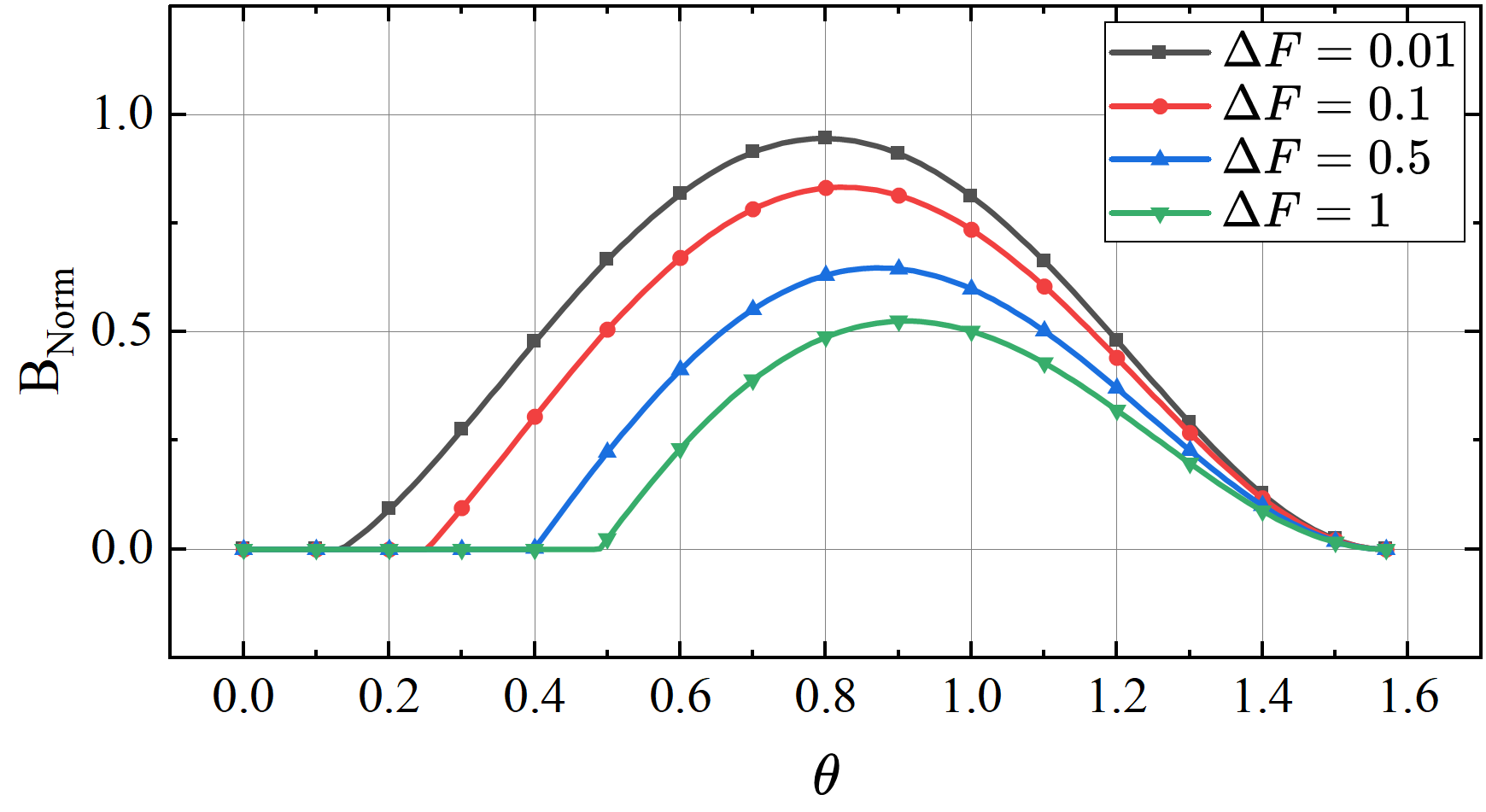}
      \caption{{The Bell nonlocality $B_{\mathrm{Norm}}\left(\hat{\rho}_{AB}\right)$, as a function of the state parameter $\theta$ with different energy ratios $\Delta F = 0.01,0.1,0.5,1$ for a pair of detectors  in Minkowski space--time. Here, the parameters $c=d=m=1$ are set.} }
     \label{fig1}
\end{figure}
\begin{figure}[h]
    \centering
   \includegraphics[width=1\linewidth]{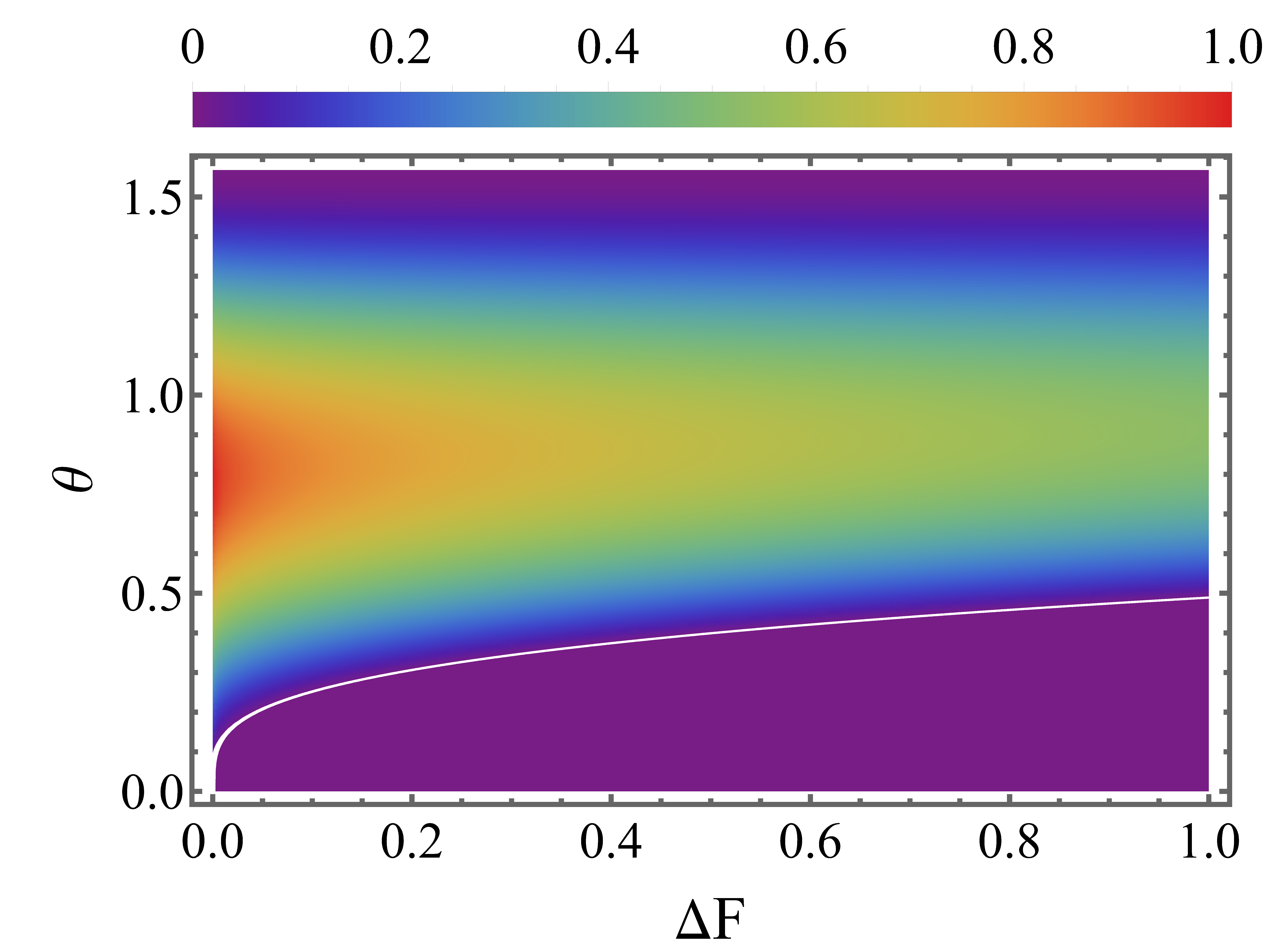}
      \caption{{The Bell nonlocality $B_{\mathrm{Norm}}\left(\hat{\rho}_{AB}\right)$ as a function of the state parameter $\theta$ and the energy ratio $\Delta F$. The white line represents the boundary distinguishing states with and without Bell nonlocality. Specifically, the regions below the white line correspond to states that do not exhibit quantum nonlocality. Here, the parameters $c=d=m=1$ are set.}}
     \label{fig2}
\end{figure}

Bell {introduced} the concept of Bell's inequality {to analyze} local realism and hidden variables \cite{PhysicsPhysiqueFizika.1.195}. {A quantum system comprising of Alice and Bob} can be represented as $\frac{1}{\sqrt{2}}\left(|01\rangle-|10\rangle\right)$ {where measurements are performed using} the bases $x\left(y\right)$. The {outcomes} of these measurements follow a conditional probability distribution, $p\left(ab|xy\right)\neq p\left(a|x\right) p\left(b|y\right)$, {indicating} that the measurement results {are} not independent. {Consequently}, if we consider all variables that would make the inequality false, the conditional probability distribution can be written as:
\begin{align}
    p\left(ab|xy,\lambda\right) = p\left(a|x,\lambda\right) p\left(b|y,\lambda\right),
    \label{E12}
\end{align}
{where} $\lambda$ {represents} a set of hidden variables {associated with each measurement, satisfying the normalization condition} $\int d\lambda q\left(\lambda\right) \equiv 1$. {Consequently}, the conditional probability distribution {can be} expressed as:
\begin{align}
    p\left(ab|xy,\lambda\right) = \int d\lambda q\left(\lambda\right) p\left(a|x,\lambda\right) p\left(b|y,\lambda\right).
    \label{E13}
\end{align}

{In} a Bell experiment, Alice and Bob {perform} different measurements, $x,y\in\left\{0,1\right\}$, and the {corresponding} outcomes are $a,b\in\left\{-1,+1\right\}$, so the expectation value of the measurements $x$ and $y$ {are} defined as {follows}:
\begin{align}
    \langle a_{x} b_{y} \rangle = \sum_{a,b} a b p\left(ab|xy\right).
    \label{E14}
\end{align}
{Further}, consider the quantity ${S} = \langle a_{0} b_{0}\rangle + \langle a_{0} b_{1} \rangle + \langle a_{1} b_{0} \rangle - \langle a_{1} b_{1} \rangle$, which is a function of $p \left(ab|xy\right)$.

If these conditional probability distributions {adhere to} local {constraints}, the quantity ${S}$ can be restrained in
\begin{align}
    {S} = \langle a_{0} b_{0}\rangle + \langle a_{0} b_{1} \rangle + \langle a_{1} b_{0} \rangle - \langle a_{1} b_{1} \rangle \le 2.
    \label{E15}
\end{align}
{This} is the {well-known} Clauser-Horne-Shimony-Holt (CHSH) inequality.

{The maximum violation of the} CHSH inequality {serves as} an effective {measure} of Bell nonlocality; for a two qubit quantum system $\hat{\rho}_{AB}$, the maximum CHSH inequality violation can be written as
\begin{align}
    B_{\mathrm{max}} \left(\hat{\rho}_{AB}\right) = 2 \sqrt{M_{AB}},
    \label{E16}
\end{align}
where $M_{AB} = \max_{i<j} \left(m_{i}+m_{j}\right)$, $m_{i(j)}(i,j=1,2,3)$ {is} the eigenvalue of matrix $\hat{T}^{\dagger}\hat{T}$, and $\hat{T}$ {is} the correlation matrix. {The} analytical {expressions} $ B_{\mathrm{max}} \left(\hat{\rho}_{AB}\right)$ {are presented} in Appendix \ref{A1}.

{Fig. \ref{fig1} depicts the variation of Bell nonlocality {with respect to} the state parameter $\theta$. {Notably}, the Bell nonlocality has been normalized, {with its} explicit expression provided in Appendix \ref{A1}. {As observed, Bell nonlocality initially increases with} the state parameter $\theta$, {before reaching a peak and subsequently decreasing.}}

{Moreover, it is evident from Fig. \ref{fig1} that not all states exhibit nonlocality in the current model. This observation raises an important question regarding which states are nonlocal and which remain local, a distinction that holds significance for practical quantum information processing. To address this, we plot the Bell nonlocality versus both the state parameter and the energy ratio $\Delta F$ in Fig. \ref{fig2}. Specifically, a boundary of states with and without Bell nonlocality was offered.
Moreover, as the energy ratio continuously increases, the range of detectable states with nonlocality gradually diminishes.}

\subsection{{$l_{1}$-norm coherence and relative entropy of coherence}}
\begin{figure*}[t]
    \centering
    \includegraphics[width=1\linewidth]{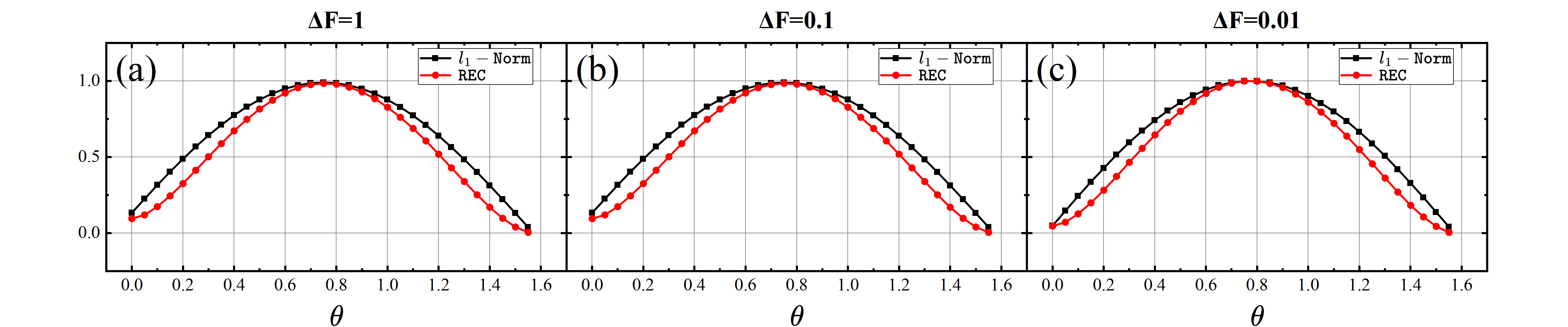}
    \caption{ {Dynamics of $l_{1}$-norm coherence (black line) and relative entropy coherence (red line) with the state parameter $\theta$ for various $\Delta F$ in Graphs (a)-(c). Graph (a): $\Delta F =1$, Graph (b): $\Delta F =0.1$, Graph (c): $\Delta F =0.01$, where $c=d=m=1$.}}
    \label{fig3}
\end{figure*}
{In quantum information science}, quantum coherence is {recognized as a fundamental quantum resource for} practical information processing. Various {quantification methods have been developed}, {including the} $l_{1}$-norm of coherence {and relative entropy of coherence}. Specifically, the $l_{1}$-norm of coherence {is widely employed due to} its operational significance and computational {feasibility}. This measure is {defined as the sum of the absolute values of} the off-diagonal elements in {the} density matrix of a quantum state, capturing the degree of coherence among the system's eigenstates. {Given a density matrix $\hat{\rho}$, the $l_{1}$-norm coherence, on the basis of eigenvectors of the Pauli spin observables $\hat{\sigma}_{i}\left(i = x,y,z\right)$, is expressed as the sum of the absolute values of its off-diagonal elements \cite{PhysRevLett.113.140401}, which can be expressed as:
\begin{align}
    C_{l_{1}}^{\hat{\sigma}_{i}}\left( \hat{\rho} \right) = \sum_{R\ne S} \langle R| \hat{\rho} | S \rangle,
    \label{E17}
\end{align}
}{where $\left\{ |R\rangle,|S\rangle \right\}$ represent the eigenvectors of $\hat{\sigma}_{i}$. For convenience, we select the $\hat{\sigma}_{z}$ Pauli spin observable as the reference basis for our calculation.} These off-diagonal elements are closely related to {the coherence of the system}, {reflecting the presence of quantum} superposition. Consequently, {a higher} $l_{1}$-norm coherence {value indicates} a greater degree of coherence {within} the quantum system.

{Additionally, the relative entropy of coherence (REC) serves as an alternative and effective measure of quantum coherence in quantum information theory. This measure is derived from the relative entropy between quantum states, which quantifies the "distance" between two quantum states.  REC characterizes the "degree of deviation" of the target quantum state from the incoherent state: the greater deviation, the stronger coherence. Mathematically, it is expressed as follows:
\begin{align}
    C_{{\rm REC}} \left( \hat{\rho} \right) = S \left( \hat{\rho}_{\rm diag} \right) - S\left( \hat{\rho} \right),
    \label{RE1}
\end{align}
where $S \left( \hat{\rho} \right) = -\mathrm{Tr} \left( \hat{\rho} \log \hat{\rho} \right)$ denotes the von Neumann entropy, and $\rho_{\rm diag}$ is the non-coherence state with the density matrix consisting only of the diagonal elements.
}

\begin{figure}[h]
    \centering
   \includegraphics[width=0.9\linewidth]{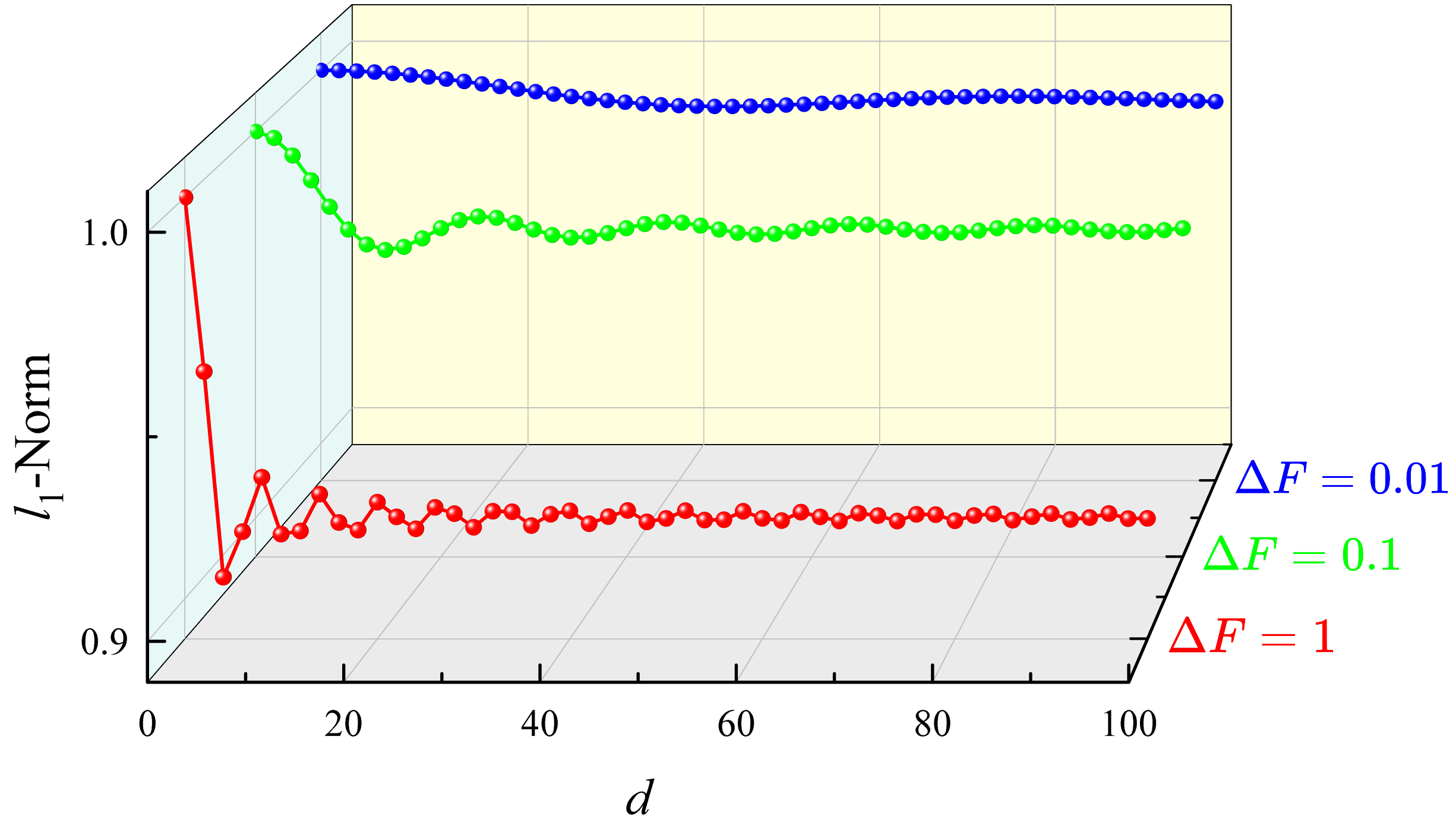}
      \caption{{Dynamics of $l_{1}$-norm coherence with the distance between two detectors. The energy ratio $\Delta F=1$ (red), $\Delta F=0.1$ (green), and $\Delta F=0.01$ (blue). $\theta = \pi/4$ is set and corresponds to the maximal entanglement, and $c=m=1$.}}
     \label{Nfig4}
\end{figure}

\begin{figure}[h]
    \centering
   \includegraphics[width=0.9\linewidth]{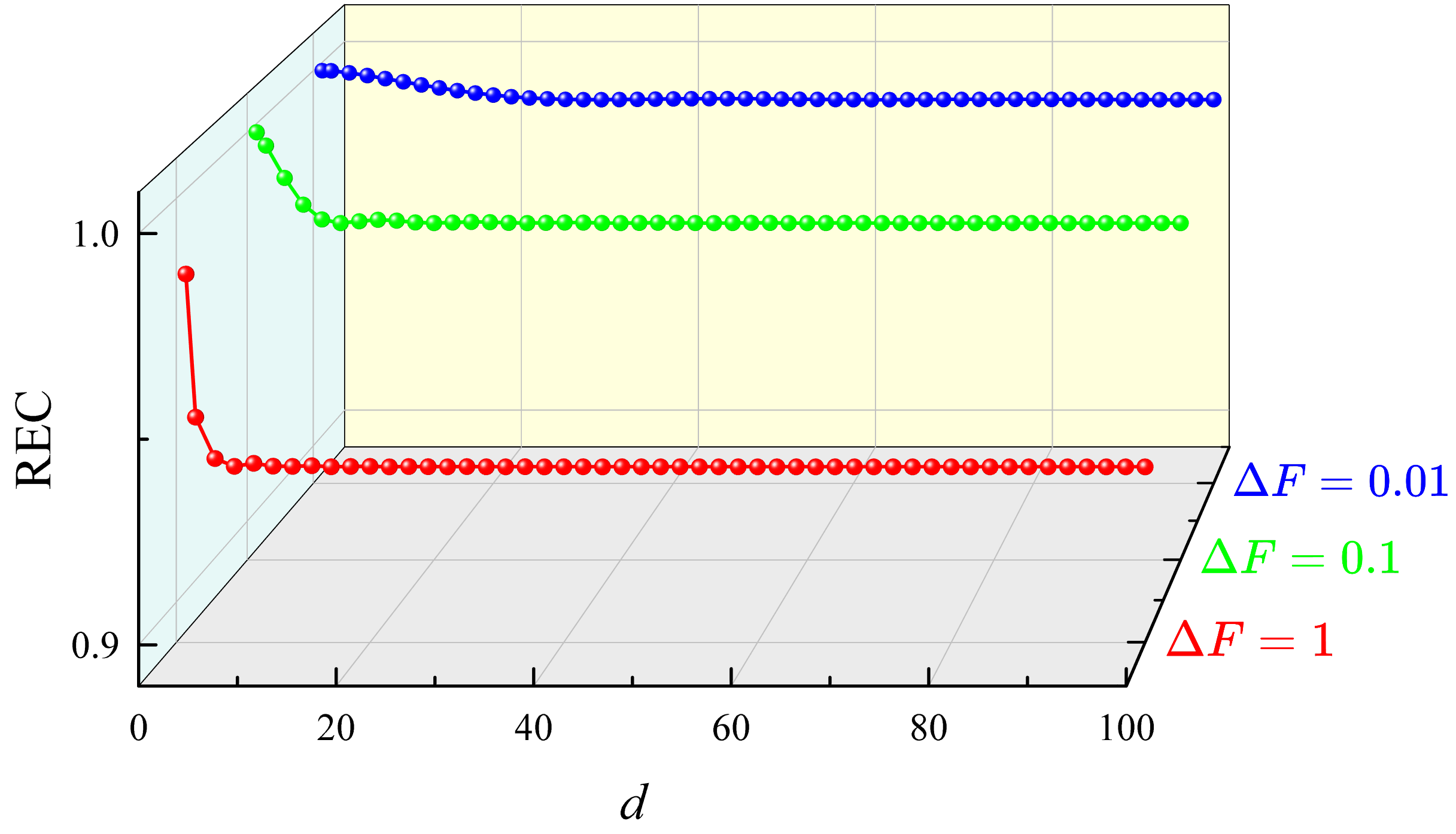}
      \caption{{Dynamics of relative entropy coherence with the distance $d$ between two detectors. The energy ratio $\Delta F=1$ (red), $\Delta F=0.1$ (green), and $\Delta F=0.01$ (blue). $\theta = \pi/4$ is set and corresponds to the maximal entanglement, and $c=m=1$.}}
     \label{Nfig5}
\end{figure}

{To analyze coherence in the present framework, Fig. \ref{fig3} illustrates the variation of the $l_{1}$-norm coherence and REC with respect to the state parameter $\theta$ for different energy ratios. Several key observations can be made: (1) Both coherence measures initially increase and then decrease as the growing state parameter grows. (2) The maximum coherence for both measures occurs at $\theta=\pi/4$. (3) The $l_{1}$-norm coherence consistently exceeds REC in this study. (4) The evolution of both coherence measures exhibits symmetry around $\theta=\pi/4$, particularly when the energy ratio is relatively small.}

{Next, we examine how the distance between the two detectors influences the system's coherence. Figs. \ref{Nfig4} and \ref{Nfig5} illustrate the dynamics of the $l_{1}$-norm coherence and REC of the two detectors with respect to the distance between the two detectors for different energy ratios, considering the detectors are initially in a maximally entangled state. The key findings are as follows: (1) Both coherence measures decrease as the distance $d$ increases, indicating that the distance can partially degrade the system's quantumness. (2) The two coherences will be frozen to fixed non-zero values for $d\rightarrow \infty$, confirming the persistence of nonlocality regardless of spatial separation between the two detectors. This behavior can be understood by analyzing the explicit expressions of the eigenstates of the density matrix, provided in Appendix \ref{A2}, which remain independent of distance $d$.}

\subsection{{Nonlocal advantage of quantum coherence}}

{Now, we briefly introduce the concept of the nonlocal advantage of quantum coherence (NAQC) based on the $l_{1}$-norm coherence and relative entropy of coherence (REC). In the previous section, we discuss the $l_{1}$-norm coherence, as defined in Eq. (\ref{E17}). However, coherence is subject to an upper bound. For mutually unbiased bases, the complete complementarity relation of coherence should satisfy the following relation:
\begin{align}
\sum_{i=x,y,z} C_{l_{1}}^{\hat{\sigma}_{i}} \le C_{\mathrm{max}},
\label{RE2}
\end{align}
which is a complementarity relation \cite{PhysRevA.95.010301}, where $C_{\mathrm{max}} = \sqrt{6} \approx 2.45$ is the upper bound, independent of the quantum state. Equality in Eq. (\ref{RE2}) is achieved for a specific pure state, $\hat{\rho}_{\max} = \frac{1}{2}\left[\frac{1}{\sqrt{3}}\left(\hat{\sigma}_{x} + \hat{\sigma}_{y} + \hat{\sigma}_{z} \right) + \hat{\mathbb{I}}\right]$, where $\hat{\mathbb{I}}$ represents the identity matrix.}

{To gain deeper insights into NAQC, one can conceptualize a game between Alice and Bob designed to demonstrate (NAQC). In this scenario, Alice and Bob each possesses qubits $A$ and $B$, respectively, with the overall quantum state represented by $\hat{\rho}_{AB}$. At the start of the game, Alice randomly performs a measurement operation $\Pi_{i}^{b} = \left[\mathbb{I} + \left(-1\right)^{b} \hat{\sigma}_{i}\right]/2$ on qubit $A$, where the value of $b$ is zero or one. The probability of obtaining a given outcome from the measurement is given by $P_{\Pi_{i}^{b}} = \mathrm{Tr}\left[\left(\Pi_{i}^{b} \otimes \mathbb{I}\right) \hat{\rho}_{AB}\right]$. Upon completing the measurement and recording the result, Alice must communicate her measurement choice and outcome to Bob. Using this information, Bob's task is to randomly measure the coherence of qubit $B$ in the eigenbases of the two remaining Pauli operators, $\hat{\sigma}_{j}$ and $\hat{\sigma}_{k}$, excluding the operator $\hat{\sigma}_{i}$ that was chosen by Alice. To determine whether the game successfully demonstrates NAQC for qubit $B$, a specific criterion must be met. Specifically, the following quantity must be calculated:
 \begin{align}
 \mathcal{N}^{l_{1}}\left(\hat{\rho}_{AB}\right) = \frac{1}{2}\sum_{i,j,k}P\left( \hat{\rho}_{\Pi_{j \ne i}}^{b} C_{l_{1}}^{\hat{\sigma} _{i}} \left( \hat{\rho}_{B|\Pi_{j\ne i}^b} \right) \right)
 \label{RE3}
 \end{align}
with $l_{1}$-norm coherence, which uses an averaging method for all possible probabilities. If  $ \mathcal{N}^{l_{1}}\left(\hat{\rho}_{AB}\right)> C_{\mathrm{max}}$ holds, then we say that NAQC for qubit $B$ is achieved.}

{Additionally, we also consider another form of nonlocal advantage based on the relative entropy of coherence.
In this case, the complete complementarity relation and the nonlocal advantage through REC are given by \cite{PhysRevA.95.010301}
\begin{align}
\sum_{i=x,y,z} C_{{\rm REC}}^{\hat{\sigma}_{i}} &\le C_{2}^m,\\
 \mathcal{N}^{{\rm REC}}\left(\hat\rho_{AB}\right) &= \frac{1}{2}\sum_{i,j,k}P\left( \hat{\rho}_{\Pi_{j \ne i}}^{b} C_{{\rm REC}}^{\hat{\sigma}_{i}} \left( \hat{\rho}_{B|\Pi_{j\ne i}^b} \right) \right).
 \label{RE5}
 \end{align}
where $C_{2}^m \approx 2.23$ represents the upper bound on the complementarity relation. Thus, the nonlocal advantage of REC is achieved when $\mathcal{N}^{{\rm REC}}\left(\hat\rho_{AB}\right)> C_{2}^m$.}

\begin{figure}[h]
    \centering
   \includegraphics[width=1\linewidth]{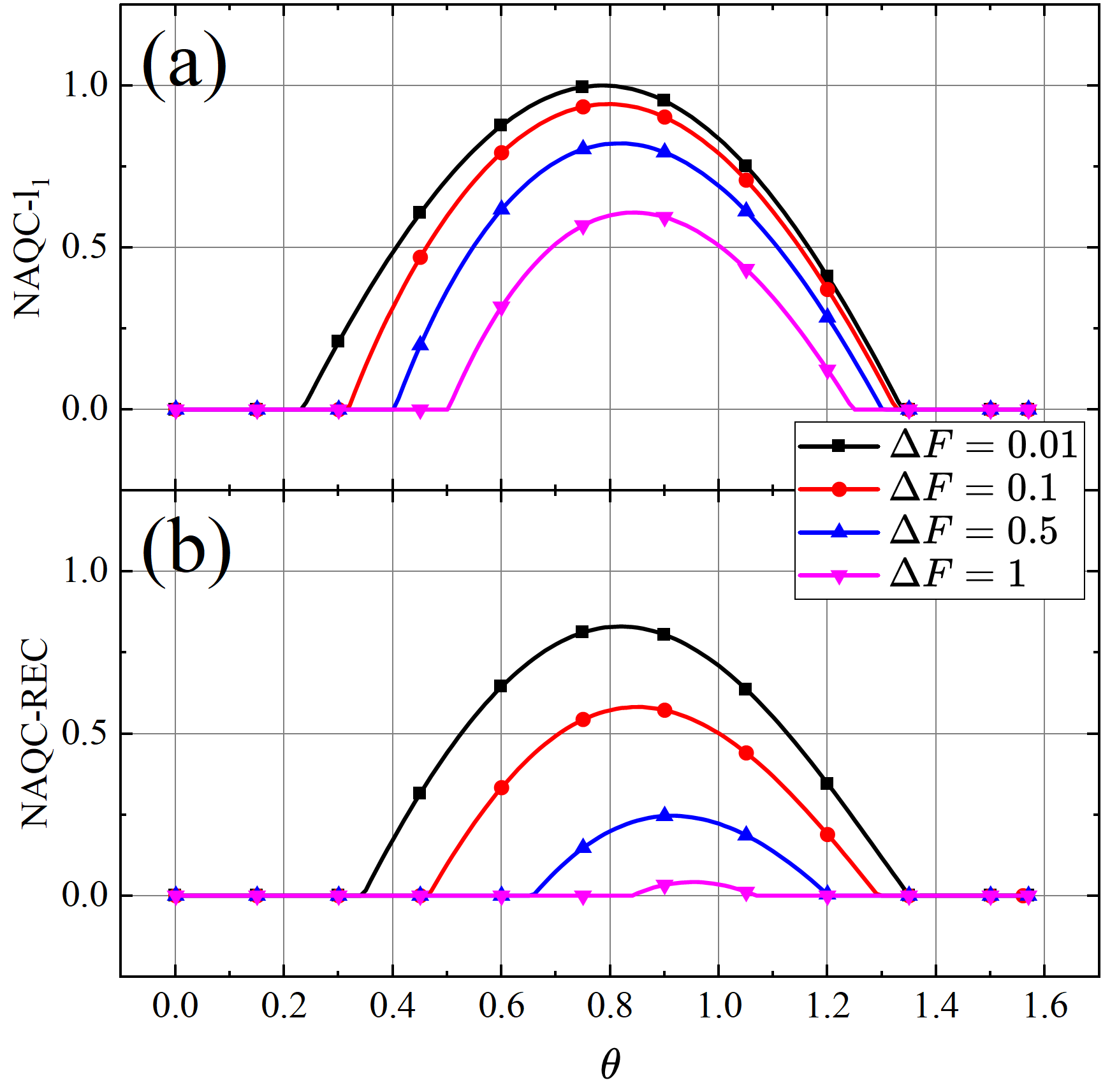}
      \caption{{Dynamics of the nonlocal advantage of quantum coherence with state parameter $\theta$ for a pair of detectors in Minkowski space--time. Graph (a): nonlocal advantage of $l_{1}$-norm coherence. Graph (b): nonlocal advantage of relative entropy coherence. The parameters are set as $c=d=m=1$ in the plots.}}
     \label{Nfig6}
\end{figure}

{Here, the analytical expressions for the nonlocal advantage of quantum coherence are provided in Appendix \ref{A3}. Additionally, for convenience, we define the normalized nonlocal advantage based on the $l_{1}$-norm coherence and REC. To further explore the characteristics of NAQC, we plot the normalized nonlocal advantage of $l_{1}$-norm coherence and REC with respect to the state parameter $\theta$ in Fig. \ref{Nfig6}. From our analysis, we conclude the following: (1) the existence of zero-valued normalized NAQC, indicating that not all states exhibit the nonlocal advantage of quantum coherence in the current model. (2) A smaller energy ratio $\Delta F$ between the detector and field enhances the nonlocal advantage of quantum coherence, suggesting that a smaller gap between them is beneficial for maximizing the system's quantumness. (3) In general, the nonlocal advantage of $l_{1}$-norm coherence exceeds that of REC, indicating that NAQC based on $l_{1}$-norm coherence is more robust than that based on REC.}

\begin{figure}[h]
    \centering
   \includegraphics[width=0.9\linewidth]{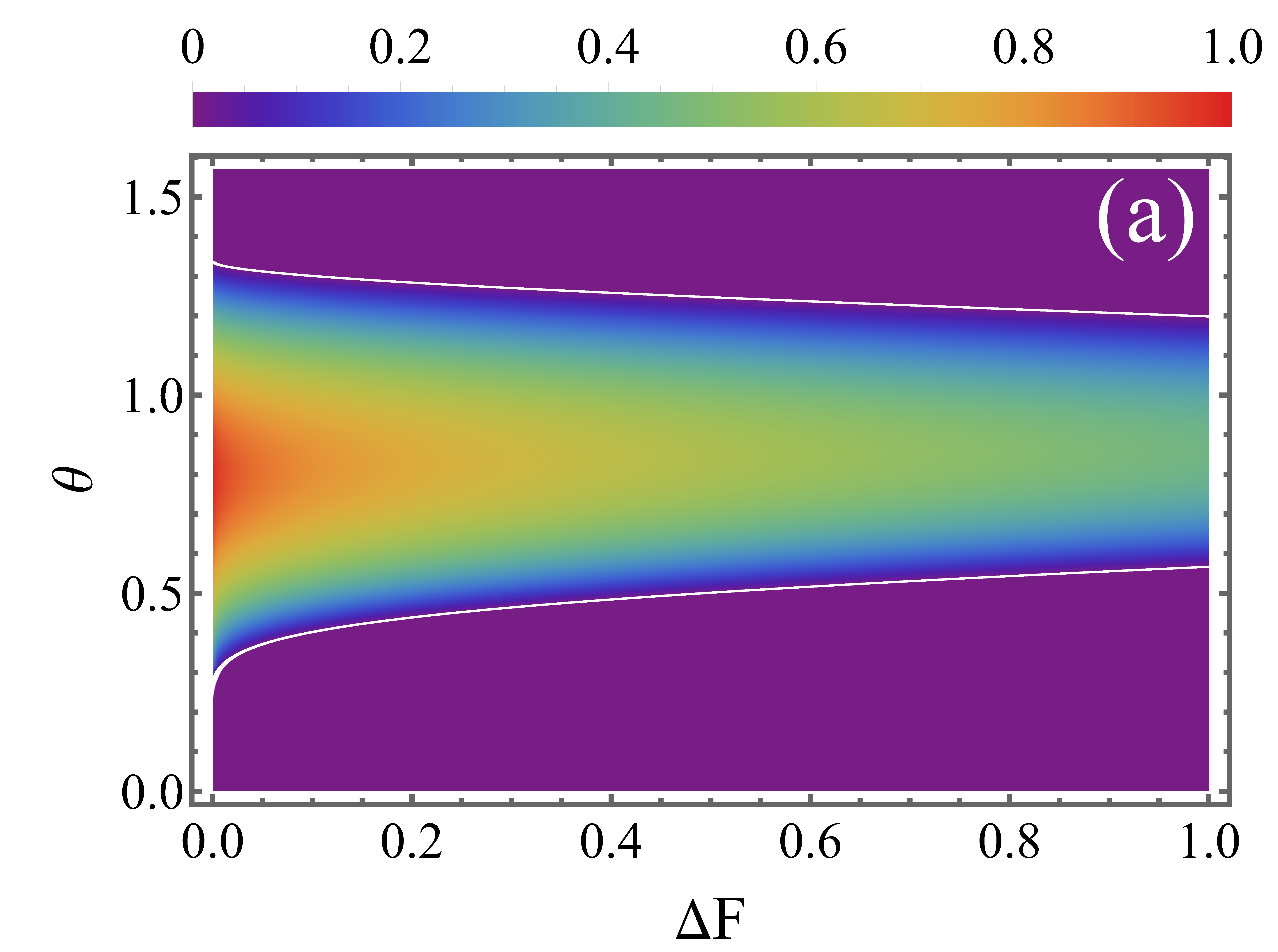}
   \centering
   \includegraphics[width=0.9\linewidth]{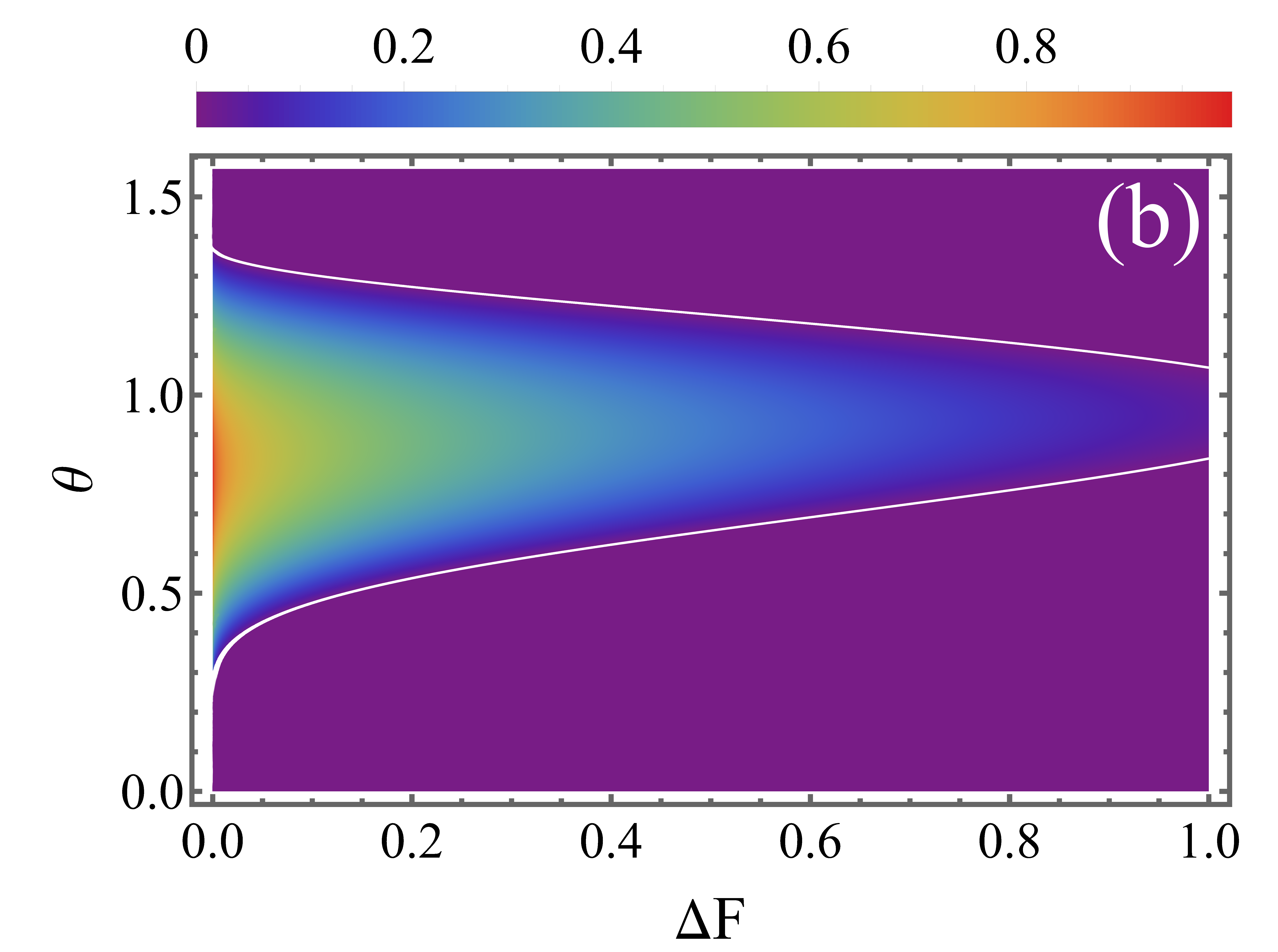}
      \caption{{Detectable state parameter range of nonlocal advantage of quantum coherence with change in energy ratio. Graph (a): Nonlocal advantage of $l_{1}$-norm coherence. Graph (b): Nonlocal advantage of relative entropy of coherence. The white lines in both graphs signify the boundaries between states with and without NAQC.  The parameters $c=d=m=1$ are set.}}
     \label{Nfig7}
\end{figure}

{Next, we investigate the boundary of the states exhibiting NAQC by plotting the nonlocal advantages of the two types of coherence with respect to the state parameter $\theta$ in Fig. \ref{Nfig7}. As illustrated in the figure, the ranges of states exhibiting NAQC gradually shrinks as the energy ratio $\Delta F$ increases, and the maximum values of the nonlocal advantage decrease accordingly. Notably, compared to the nonlocal advantage of REC, the detectable range of the nonlocal advantage of $l_{1}$-norm coherence is broader and less affected by the energy ratio. Thus, we argue that the nonlocal advantage of $l_{1}$-norm coherence is more robust than that of REC, which aligns with our previous conclusions.}

\subsection{Entropic uncertainty and lower bound}

\begin{figure*}[t]
    \centering
    \includegraphics[width=1\linewidth]{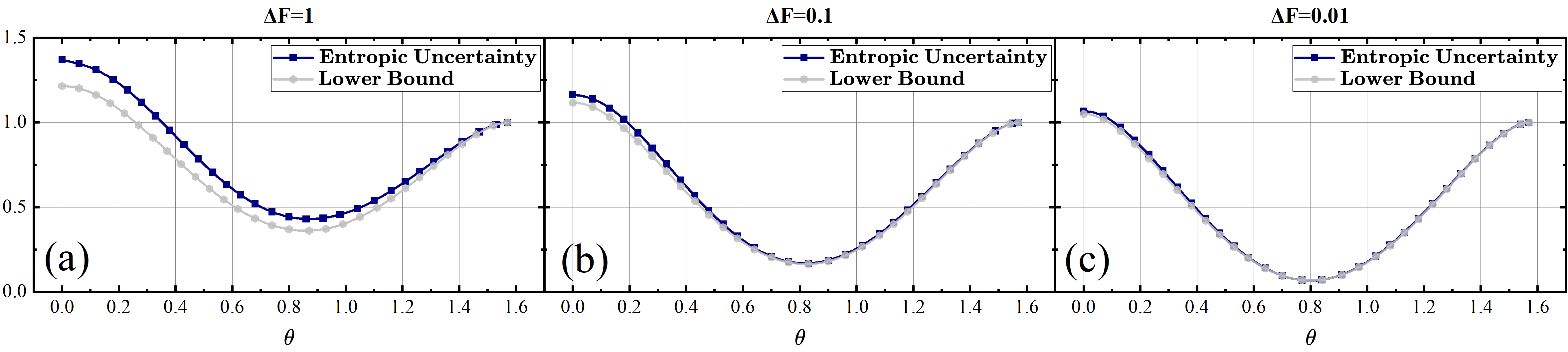}
    \caption{ {Uncertainty and lower bound with the state's parameter $\theta$ for various $\Delta F$ in Graphs (a)--(c). Graph (a): $\Delta F=1$, Graph (b): $\Delta F=0.1$, and Graph (c): $\Delta F=0.01$, where $c=d=m=1$.}}
    \label{Nfig8}
\end{figure*}

{The entropic} uncertainty relation (EUR) was {first} formulated by Deutsch \cite{PhysRevLett.50.631}, {later refined} by Karus \cite{PhysRevD.35.3070}, and {rigorously} by Maassen and Uffink \cite{PhysRevLett.60.1103}. {It is} expressed as
\begin{equation}
    \begin{aligned}
        H(\hat{R}) + H(\hat{S}) \ge \log_{2}\frac{1}{c} := q_{MU},
        \label{E19}
    \end{aligned}
\end{equation}
{where} $H\left({X}\right) = -\Sigma_{k} x_{k} \log_{2} x_{k}$ {denotes} the Shannon entropy of {the} observable ${X} \in \left\{\hat{R},\hat{S}\right\} $, $x_{k}$ {denotes} the probability of {obtaining} outcome $k$, and $q_{MU}$ denotes the incompatibility measure, {defined as} $c = \max_{i,j} |\langle \hat{r}_{i}|\hat{s}_{j}\rangle|^{2}$; here, $|\hat{r}_{i}\rangle$ and $|\hat{s}_{j}\rangle$ are the eigenstates of {$\hat{R}$ and $\hat{S}$}, respectively. For {composite quantum systems}, Renes $et\ al$ and Berta $et\ al$ {proposed} quantum-memory-assisted entropic uncertainty relations (QMA-EUR) {for} arbitrary {pairs of} observables, which can be mathematically {expressed as follows}:
\begin{align}
    S(\hat{R}|{B})+S(\hat{S}|{B})\ge q_{MU}+S({A}|{B}),
    \label{E20}
\end{align}
where $S(\hat{R}|{B})=S(\hat{\rho}_{\hat{R}B})-S(\hat{\rho}_{B})$ and $S(\hat{S}|{B})=S(\hat{\rho}_{\hat{S}B})-S(\hat{\rho}_{B})$ {represent the} von Neumann entropies of the post-measurement states. {Consequently}, the quantum states can {be expressed as}:
\begin{align}
    \begin{split}
        \hat\rho_{\hat{X}B} & = \sum_{i}\left(|{x}_{i}\rangle \langle {x}_{i}|\otimes\mathbb{\hat{I}}\right) \hat{\rho}_{AB} \left(|{x}_{i}\rangle \langle {x}_{i}| \otimes \mathbb{\hat{I}}\right), \\
        \hat\rho_{\hat{Z}B} & = \sum_{i}\left(|{z}_{i}\rangle \langle {z}_{i}|\otimes\mathbb{\hat{I}}\right) \hat{\rho}_{AB} \left(|{z}_{i}\rangle \langle {z}_{i}| \otimes \mathbb{\hat{I}}\right),
        \label{E21}
    \end{split}
\end{align}
after performing two Pauli measurements, $\hat{X}$ and $\hat{Z}$, where $\mathbb{\hat{I}}$ is the identity matrix, {and} $|{x}_{i}\rangle$ and $|{z}_{i}\rangle$ are the eigenvectors of the corresponding Pauli matrices. {Significant} progress {has} {been made in the study of QMA-EUR}. To {better} understand QMA-EUR, we {consider} an uncertainty game between two legitimate players, Alice and Bob. {The results obtained were in agreement}. {In} two measurements, $\hat{R}$ and $\hat{S}$, one of the players, say Bob, prepares two particles, ${A}$ and ${B}$ in {an} entangled state. Then, Bob sends particle ${A}$ to {Player} Alice {while keeping} $ {B}$ as quantum memory. {Subsequently}, Alice {selects} either $\hat{R}$ or $\hat{S}$ {for measurement and records} the outcome. {She then communicates her measurement choice to} Bob via {a} classical channel. Bob's {task} is to {predict Alice's} result with minimal uncertainty, {which is constrained} by Eq. (\ref{E20}). {To be explicit, the entropic uncertainty and its lower bound in the current model have been derived, as detailed in Appendix \ref{A4}.}

Fig. \ref{Nfig8}(a)-(c) depicts the entropic uncertainty and its lower bound with $\theta$ for different energy ratios $\Delta F$. {As observed in} the figures, the entropic uncertainty {initially} decreases and then increases to a fixed value with increasing parameter $\theta$. {Notably}, the uncertainty is {strongly anticorrelated} with quantum coherence, {as shown in} Fig. \ref{fig3}. {Interestingly, the analysis reveals that} a smaller energy ratio $\Delta F$ {results in a tighter bound, meaning that} the difference between the uncertainty and its lower bound {decreases}. {This suggests that, for relatively small $\Delta F$}, the bound {effectively captures} the uncertainty, {as further illustrated} in Fig. \ref{Nfig8}. {Additionally}, the decline in quantum resources mentioned above can be attributed to the interaction {between} the detector and the environment, which reduces the system's purity. From {an} entropy {perspective}, the loss of {information within the} system {increases its} uncertainty, {implying that less information can be extracted} from the system of interest. Consequently, {this study provides a} promising {new} perspective {for quantifying} the quantumness of the system.

\begin{figure}[t]
    \centering
    \includegraphics[width=1\linewidth]{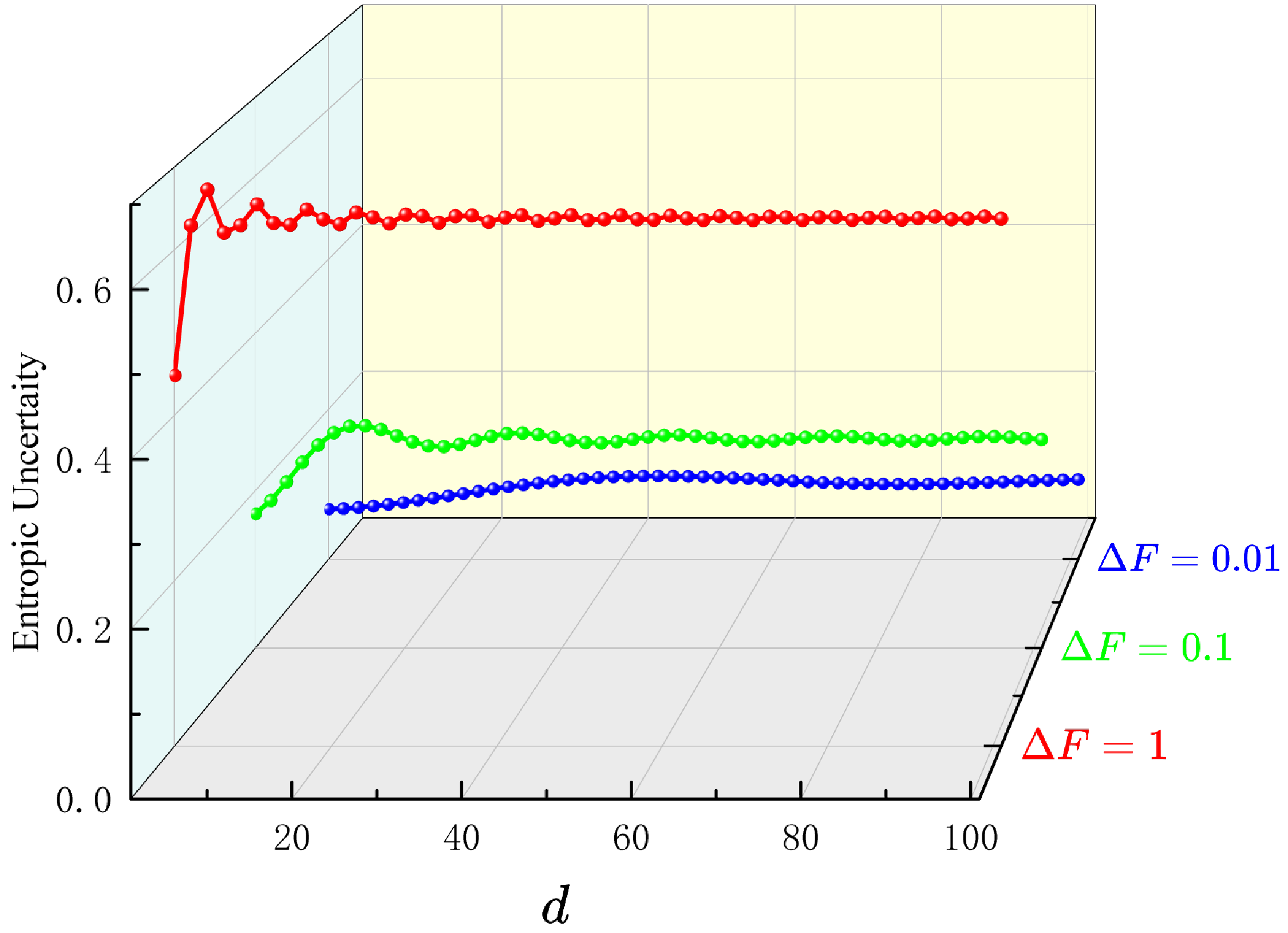}
    \caption{{Uncertainty with the distance between two detectors, with the energy ratio set as $\Delta F=1$ (red), $\Delta F=0.1$ (green), and $\Delta F=0.01$ (blue), where $c=m=1$.}}
    \label{Nfig9}
\end{figure}
{Finally, we explore how the distance $d$ between the two detectors influences the evolution of entropic uncertainty.
As depicted in Fig. \ref{Nfig9}, the uncertainty exhibits oscillatory growth as the distance $d$ increases, and eventually stabilizes at a fixed value when the detectors are sufficiently far apart, i.e., $d\rightarrow\infty$. This behavior can be attributed to the minimal quantumness of the system in this regime, as shown in Figs. \ref{Nfig4}  and \ref{Nfig5}, which results in maximal measurement uncertainty.}

\section{Conclusion}
{In this work, we explored the dynamics of quantum nonlocality, quantum coherence, the nonlocal advantage of quantum coherence, and measurement uncertainty within a relativistic framework. Specifically, we considered a pair of static Unruh-DeWitt detectors interacting independently with an external field. Our analysis revealed how state parameters, the energy ratio, and the distance between the detectors influence the quantumness of the system. Notably, several meaningful results were obtained: (i) A larger energy ratio reduces both Bell nonlocality and quantum coherence in the Unruh-DeWitt detector system, with higher energy ratios $\Delta F$ degrading the system's quantumness. (ii) An increasing distance $d$ between the detectors weakens both the $l_{1}$-norm coherence and the relative entropy of  coherence, eventually stabilizing at fixed values as the distance approaches infinity. (iii) Regarding the nonlocal advantage of coherence, the $l_{1}$-norm coherence is more robust than the relative entropy of coherence within the relativistic framework. (iv) The measured uncertainty of the system is strongly correlated with nonlocality and quantum coherence. Overall, these findings provide deeper insights into the quantumness of a pair of static Unruh-DeWitt detectors and hold fundamental significance for future quantum information processing in relativistic settings.}

\begin{acknowledgements}
    This work was supported by the National Natural Science Foundation of China (Grant Nos. 12475009 and 12075001, and 62471001), Anhui Provincial Key Research and
Development Plan (Grant No. 2022b13020004), Anhui Province Science and Technology Innovation Project
(Grant No. 202423r06050004), and Anhui Provincial University Scientific Research Major Project (Grant No.
2024AH040008).
\end{acknowledgements}

\bibliographystyle{plain}


\appendix
\begin{widetext}
\section{{Analytical} results of Bell nonlocality}\label{A1}
In the {calculation above}, we {derived} the final state $\hat{\rho}_{AB}$ of the {entire} system. {Subsequently}, by inserting matrix into Eq. (\ref{E16}), the maximum CHSH inequality violation can be {explicitly expressed} as:
\begin{align}
  B\left(\hat{\rho}_{AB}\right) = 2 \sqrt{\frac{c \pi (\alpha^{2} + \gamma^{2}) - \gamma^{2} \mathrm{Re}\left(\sqrt{\Delta F}\right)}{c^{2}\pi^{2}}+\frac{\gamma^2 \left( -4 c d \pi \alpha + d \alpha \mathrm{Re}\left(\sqrt{\Delta F}\right) + c \gamma \sin \left(\frac{d \mathrm{Re}\left(\sqrt{\Delta F}\right)}{c}\right)\right)}{4 c^2 d^2 \pi ^2}}.
\end{align}
{To enhance the interpretability of our results, we perform a normalization procedure on the above formula,
\begin{align}
  B_{\mathrm{Norm}}\left(\hat{\rho}_{AB}\right) := \max\left\{0, \frac{B\left(\hat{\rho}_{AB}\right)-2}{B_{{\mathrm{max}}}\left(\hat{\rho}_{AB}\right)-2}\right\},
\end{align}
where $B_{\mathrm{max}}\left(\hat{\rho}_{AB}\right) = 2\sqrt{2}$.}
{Here}, $\alpha = \sin\theta$ and $\gamma = \cos\theta$ are state parameters. Referring to Eq. (\ref{E9}) because the detectors {are same}, we {obtain}
$P = P^{''}_{A}/T = P^{''}_{B}/T = \frac{\delta\left(0\right)}{2c^{3}T} \sqrt{\Delta F}$, and $\dot{\Re}_{AB} = \Re_{AB}/T= \frac{\delta\left(0\right)}{2c^{3}/T} \sqrt{\Delta F} \frac{\sin\left(\frac{d}{c} \sqrt{\Delta F}\right)}{\left(\frac{d}{c} \sqrt{\Delta F}\right)}$.

\section{{Analytical} results of $l_{1}$-norm coherence and relative entropy of coherence}\label{A2}
Resorting to Eq. (\ref{E17}), the $l_{1}$-norm coherence {is given by}
\begin{align}
 C_{l_{1}} (\hat{\rho}_{AB}) & = 2 \alpha \gamma \left(1 - 2c^2 M\right) + 2 \gamma^2 c^2 \dot{\Re}_{AB},
\end{align}
where $M = \mathrm{Re}(M_{A})/T = \mathrm{Re}(M_{B})/T = \frac{\delta\left(0\right)}{4c^{3}T} \sqrt{\Delta F}$. {When distance $d \rightarrow \infty$, the results of $l_{1}$-norm coherence can be written as $C_{l_{1}}^{(d \rightarrow \infty)} (\hat{\rho}_{AB}) \simeq 2 \alpha \gamma \left(1 - 2c^2 M\right)$, is independent of the distance.}

{Resorting to Eq. (\ref{RE1}), $\rho$ is the final state; we consider $\hat{\rho}_{AB}(t)$ and $\hat{\rho}_{\rm diag}$ as the incoherent states of $\hat{\rho}_{AB}(t)$, which can be written as follows:
\begin{align}
 \hat{\rho}_{AB}(t)                    & = \begin{pmatrix}
                                      \rho_{11} & 0         & 0         & \rho_{14} \\
                                      0         & \rho_{22} & \rho_{23} & 0         \\
                                      0         & \rho_{32} & \rho_{33} & 0         \\
                                      \rho_{41} & 0         & 0         & \rho_{44}
                                    \end{pmatrix},\\
 \hat{\rho}_{\rm diag}                & = \begin{pmatrix}
                                      \rho_{11} &        0         &       0   &    0 \\
                                      0         &    \rho_{22}     &       0   &    0 \\
                                      0         &        0         & \rho_{33} &    0 \\
                                      0         &        0         &       0   & \rho_{44}
                                    \end{pmatrix}.
\end{align}
According to the von Neumann entropy $S \left( \hat{\rho} \right) = -\mathrm{Tr} \left( \hat{\rho} \log \hat{\rho} \right)$, the relative entropy of the coherence can be written as
\begin{align}
C_{\rm REC}(\hat{\rho}_{AB}) = -\sum_{i=1,2,3,4} \lambda_{i} \log_{2}(\lambda_{i}) + \sum_{k=1,2,3,4} \mu_{k} \log_{2}(\mu_{k}),
\end{align}
where $\lambda_{i}(i=1,2,3,4)$ are the eigenvalues of $\hat{\rho}$ and $\mu_{k}(k=1,2,3,4)$ are the eigenvalues of $\hat{\rho}_{\rm diag}$; in detail, $\lambda_{1} = \frac{\Re - \aleph}{4c^2d\pi}$, $\lambda_{2}=\frac{\Re + \aleph}{4c^2d\pi}$,
$\lambda_{3}=\frac{\cos^2(\theta)\left(d \mathrm{Re} \left(\sqrt{\Delta F}\right)-c\sin\left(d \mathrm{Re} \left(\sqrt{\Delta F}\right)/c\right) \right)}{4cd\pi}$
$\lambda_{4}=\frac{\cos^2(\theta)\left(d \mathrm{Re} \left(\sqrt{\Delta F}\right)+c\sin\left(d \mathrm{Re} \left(\sqrt{\Delta F}\right)/c\right) \right)}{4cd\pi}$
\begin{align}
\Re := 2 c^2 d \pi - c d \cos^2\left(\theta\right) \mathrm{Re} \left( \sqrt{\Delta F} \right), \nonumber
\aleph := \sqrt{c^2 d^2 \left( 4 c^2 \pi^2 + \cos^2(\theta) \mathrm{Re}(\sqrt{\Delta F}) \left(-4 c \pi +\mathrm{Re}(\sqrt{\Delta F})\right) \right)}.
\end{align}
Moreover, when the distance $d \rightarrow \infty$, the eigenvalues are calculated as
\begin{align}
\lambda_{1}(d \rightarrow \infty) & \simeq \frac{2 c \pi - \cos^2(\theta) {\rm Re}\sqrt{\Delta F}-\sqrt{( 4 c^2 \pi^2 + \cos^2(\theta) \mathrm{Re}(\sqrt{\Delta F})(-4 c \pi +\mathrm{Re}(\sqrt{\Delta F})))}}{4 \pi}, \\ \nonumber
\lambda_{2}(d \rightarrow \infty) & \simeq \frac{2 c \pi - \cos^2(\theta) {\rm Re}\sqrt{\Delta F}+\sqrt{( 4 c^2 \pi^2 + \cos^2(\theta) \mathrm{Re}(\sqrt{\Delta F})(-4 c \pi +\mathrm{Re}(\sqrt{\Delta F})))}}{4 \pi}, \\
\lambda_{3}(d \rightarrow \infty) & = \lambda_{4}(d \rightarrow \infty) \simeq \frac{\cos^2(\theta) {\rm Re} \sqrt{\Delta F}}{4 c \pi}. \nonumber
\end{align}
Furthermore, $\mu_{k}(k=1,2,3,4)$ are the eigenvalues of $\hat{\rho}_{\rm diag}$, and they are given by
$\mu_{1} = \frac{\cos^2(\theta) (2 c \pi - \mathrm{Re}(\sqrt{\Delta F}))}{2 c \pi}$.
$\mu_{2} = \mu_{3} = \frac{\cos^{2}(\theta) \mathrm{Re}(\sqrt{\Delta F})}{4 c \pi}$
$\mu_{4} = \sin^2(\theta)$.
}
{\section{Analytical expressions of the nonlocal advantage of quantum coherence}\label{A3}
By utilizing Eq. (\ref{RE3}), the analytical result for nonlocal advantage of $l_{1}$-norm coherence can be expressed as follows:
\begin{align}
\mathcal{N}^{l_{1}}(\hat{\rho}_{AB})
& = \frac{4 \pi c - 3 \sqrt{\Delta F} \cos^2\theta (  \sqrt{\Delta F} +4 \pi c (\sec^2\theta -1) }{ 4 \pi c(4  \pi c-\sqrt{\Delta F})} \nonumber \\
& + \frac{\cos^2\theta \left( c \cos\theta \sin(\frac{d \sqrt{\Delta F}}{c}) +d (- 4 \pi c +\sqrt{\Delta F} \sin\theta) \right)}{4 \pi c d} \nonumber \\
& + \frac{(- 4 \pi c + \sqrt{\Delta F}) \cos^2 \theta (\sqrt{\Delta F} \cos^2 \theta - 4 \pi c \sin^2 \theta  )}
         {2 c (- \pi (4 \pi c + \sqrt{\Delta F}) + \pi (4 \pi c - \sqrt{\Delta F}) \cos(2\theta))} \nonumber \\
& + \frac{\cos^2 \theta \left( c \sin(\frac{d \sqrt{\Delta F}}{c})+d (4 \pi c -\sqrt{\Delta F} )\tan\theta \right)}{4 \pi c d} \\
& + \frac{1}{2} \sqrt{\frac{\cos^2\theta \left( c \cos\theta \sin (\frac{d\sqrt{\Delta F}}{c} ) + d ( -4 \pi c + \sqrt{\Delta F}) \sin\theta \right)^2}{4 \pi^2 c^2 d^2 }
  + \left(\left(-1+\frac{\sqrt{\Delta F} }{2 \pi c } \cos^2 \theta\right)+\sin^2 \theta \right)^2 } \nonumber \\
& +\frac{1}{2} \sqrt{\frac{\cos^4\theta \left( c \sin (\frac{d\sqrt{\Delta F}}{c} ) + d ( 4 \pi c - \sqrt{\Delta F}) \tan\theta \right)^2}{4 \pi^2 c^2 d^2 }
  + \left(\left(-1+\frac{\sqrt{\Delta F} }{2 \pi c } \cos^2 \theta\right)+\sin^2 \theta \right)^2 }\nonumber.  \nonumber
\end{align}
Because the analytical results of nonlocal advantage of the REC are highly complex, we do not present them in this paper.}
{To make the results more intuitive, we perform a normalization operation on the above formula, yielding the following results:
\begin{align}
\mathcal{N}^{l_{1}}_{\mathrm{Norm}}(\hat{\rho}_{AB})& := \max \left\{0,\frac{\mathcal{N}^{l_{1}}(\hat{\rho}_{AB})-\sqrt{6}}{\mathcal{N}^{l_{1}}_{\mathrm{max}}(\hat{\rho}_{AB})-\sqrt{6}} \right\}, \\
\mathcal{N}^{\rm REC}_{\mathrm{Norm}}(\hat{\rho}_{AB})& := \max \left\{0,\frac{\mathcal{N}^{\rm REC}(\hat{\rho}_{AB})-2.23}{\mathcal{N}^{\rm REC}_{\mathrm{max}}(\hat{\rho}_{AB})-2.23} \right\},
\end{align}
where $\mathcal{N}^{l_{1}}_{\mathrm{max}} (\hat{\rho}_{AB}) = \mathcal{N}^{\rm REC}_{\mathrm{max}} (\hat{\rho}_{AB}) = 3$}.

\section{{Analytical} expressions of the entropic uncertainty and lower bound}\label{A4}
To {compute} the uncertainty and its lower bound, which correspond to {the left and right sides} of Eq. (\ref{E20}), we {consider} the {state of the system} after {measurements with the} two incompatible observables $\hat{\sigma}_{x}$ and $\hat{\sigma}_{z}$, as {follows}:
\begin{align}
\hat{\rho}_{\hat{X}B} & =
\begin{pmatrix}
  \frac{\rho_{11}+\rho_{33}}{2} &                 0              &                 0               & \frac{\rho_{14}+\rho_{32}}{2}\\
                 0              &  \frac{\rho_{22}+\rho_{44}}{2} &  \frac{\rho_{23}+\rho_{41}}{2}  &                0             \\
                 0              &  \frac{\rho_{14}+\rho_{32}}{2} &  \frac{\rho_{11}+\rho_{33}}{2}  &                0             \\
  \frac{\rho_{23}+\rho_{41}}{2} &                 0              &                 0               & \frac{\rho_{22}+\rho_{44}}{2}
\end{pmatrix},\\
\hat{\rho}_{\hat{Z}B} & =
\begin{pmatrix}
   \rho_{11} &     0     &     0     &     0    \\
       0     & \rho_{22} &     0     &     0    \\
       0     &     0     & \rho_{33} &     0    \\
       0     &     0     &     0     & \rho_{44}
\end{pmatrix}.
\end{align}
Furthermore, the reduced density matrix of Bob's detector {can be expressed as:}
\begin{align}
\hat{\rho}_{B}=
\begin{pmatrix}
 \rho_{11}+\rho_{33} &          0          \\
          0          & \rho_{22}+\rho_{44}
\end{pmatrix},
\end{align}
by {tracing out} the {degrees of freedom} of Alice's detector. {The explicit expression for} the entropic uncertainty {is then given by}
\begin{align}
S\left(\hat{X}|B\right) + S\left(\hat{Z}|B\right) = -\sum_{i}\lambda_{i} \log_{2}\left(\lambda_{i}\right) - \sum_{i}\epsilon_{i} \log_{2}\left(\epsilon_{i}\right) + 2\sum_{k}\mu_{k}\log_{2}\left(\mu_{k}\right),
\end{align}
where $
\lambda_{1} = \lambda_{2} = \frac{1}{4} (1 - \Gamma) \nonumber; \  \lambda_{3}   = \lambda_{4} = \frac{1}{4} (1 + \Gamma)$ are the eigenvalues of the matrix $\hat{\rho}_{\hat{X}B}$ with $
\Gamma = \sqrt{\alpha^4 + 2( 1 + 2 c^2 (4 M (-1 +c^2M) + P ))\alpha^2 \gamma^2 + 8 c^2 (1 - 2 c^2 M)\dot{\Re}_{AB} \alpha \gamma^3 + ((1 - 2 c^2 P)^2 + 4 c^4 \dot{\Re}_{AB}^2) \gamma^4}$,   $
\epsilon_{1}  = (1 - 2 c^2 P) \gamma^2, \ \epsilon_{2}   = \epsilon_{3} = \gamma^2 c^2 P$ and $\epsilon_{4}   = \alpha^2 $ are the eigenvalues of the matrix $\hat{\rho}_{\hat{Z}B}$, and $\mu_{1}   = (1 - c^2P) \gamma^2$ and $\mu_{2} = \alpha + c^2 P \gamma^2$ are the eigenvalues of  $\hat{\rho}_{B}$.

{Additionally}, the lower bound of {uncertainty} can be written as:
\begin{align}
q_{MU} + S\left(A|B\right) = 1 + \left( - \sum_{i=1}^4 \zeta_{i} \log_2\left( \zeta_{i} \right) + \sum_{k=1}^2 \mu_{k} \log_{2}\left(\mu_{k}\right)\right),
\end{align}
where $\left\{\zeta_{i}|i=1,2,3,4\right\}$ is the eigenvalue of the matrix $\hat{\rho}_{AB}$, {and it can be} expressed by
\begin{align}
\zeta_{1} & = c^2 (P - \dot{\Re}_{AB}) \gamma^2, \nonumber \\
\zeta_{2} & = c^2 (P - \dot{\Re}_{AB}) \gamma^2,\nonumber \\
\zeta_{3} & = \frac{1}{2} (\alpha^2 + (1 - 2 c^2 P) \gamma^2 - \Xi), \nonumber \\
\zeta_{4} & = \frac{1}{2} (\alpha^2 + (1 - 2 c^2 P) \gamma^2 + \Xi)  \nonumber
\end{align}
with $\Xi=\sqrt{(\alpha^4 + 2 (1 + 2 c^2 (4 M (-1 + c^2 M) + P))\alpha^2 \gamma^2 + (1 - 2 c^2 P)^2 \gamma^4)}$.

\end{widetext}

\end{document}